\documentclass[10pt,letterpaper,twocolumn]{article} 

\usepackage{ol2}
\usepackage[draft]{hyperref}
\usepackage{amsmath}

\begin{document}

\twocolumn[ 

\title{Higher-order defect mode laser\\in an optically thick photonic crystal slab}


\author{Se-Heon Kim,$^{1,3,*}$ Jingqing Huang,$^{2,3}$ and Axel Scherer$^{1,2,3}$}

\address{
$^1$Department of Physics, California Institute of Technology, Pasadena, CA 91125, USA
\\
$^2$Department of Electrical Engineering, California Institute of Technology, Pasadena, CA 91125, USA
\\
$^3$Kavli Nanoscience Institute, California Institute of Technology, Pasadena, CA 91125, USA
\\
$^*$Corresponding author: seheon@caltech.edu
}

\begin{abstract}
The use of an optically thick slab may provide versatile solutions for the realization of a current injection type laser using photonic crystals. Here, we show that a transversely higher-order defect mode can be designed to be confined by a photonic band gap in such a thick slab. Using simulations, we show that a high-$Q$ of $> 10^5$ is possible from a finely tuned second-order hexapole mode. Experimentally, we achieve optically pumped pulsed lasing at 1347 nm from the second-order hexapole mode with a peak threshold pump power of 88 $\mu$W. 
\end{abstract}

\ocis{230.5298, 250.5960.}

 ] 

Two-dimensional (2-D) photonic crystal (PhC) slab structures have, so far, been in the form of a {\em thin} dielectric slab, whose thickness $T$ is often chosen to be $\sim$200 nm for an operational wavelength of $\sim$1.3 $\mu$m. This thickness consideration is to maximize the size of the photonic bandgap (PBG) in the  in-plane direction ($x$-$y$ plane)\cite{JohnsonFa99}, which has unfortunately placed a severe constraint on the design of a current-injection type laser. Pulsed lasing operation has been demonstrated using a vertically-varying {\em p-i-n} doping structure within the thin PhC slab, for which a sub-micron size dielectric post placed directly underneath the laser cavity serves as a current path\cite{H_G_Park_Science04}. Recent efforts have moved towards a {\em laterally}-varying {\em p-i-n} structure and a few successful results were already reported by groups in both Stanford\cite{Ellis11} and NTT\cite{Matsuo_OPEX12}. However, there are still favorable reasons for using a vertically-varying doped structure, because such a design allows a monolithic growth of all of the epitaxial layers that are almost free of crystal defects. 


Recently, we have shown that even a {\em very thick} slab can support sufficiently high-$Q$ (few thousands) cavity modes for lasing.\cite{S_H_Kim_OL12} In our previous result, however, the dipole mode formed in a triangular lattice air-hole PhC slab was emitting more photons into the in-plane directions rather than into the vertical direction ($z$) for efficient photon emission and collection. Moreover, $Q$ could not exceed 3,000 with $T$ = 606 nm. It would seem, at first, that we have no other options for further improvement in $Q$, since the poor horizontal confinement appears inevitable due to the absence of a PBG. It is our purpose in this Letter to rebut this first intuition and show that the thick slab can be used to achieve an efficient vertical emitter with a surprisingly high $Q$ of over $10^5$. 

\begin{figure*}[t]
\centering\includegraphics[width=17cm]{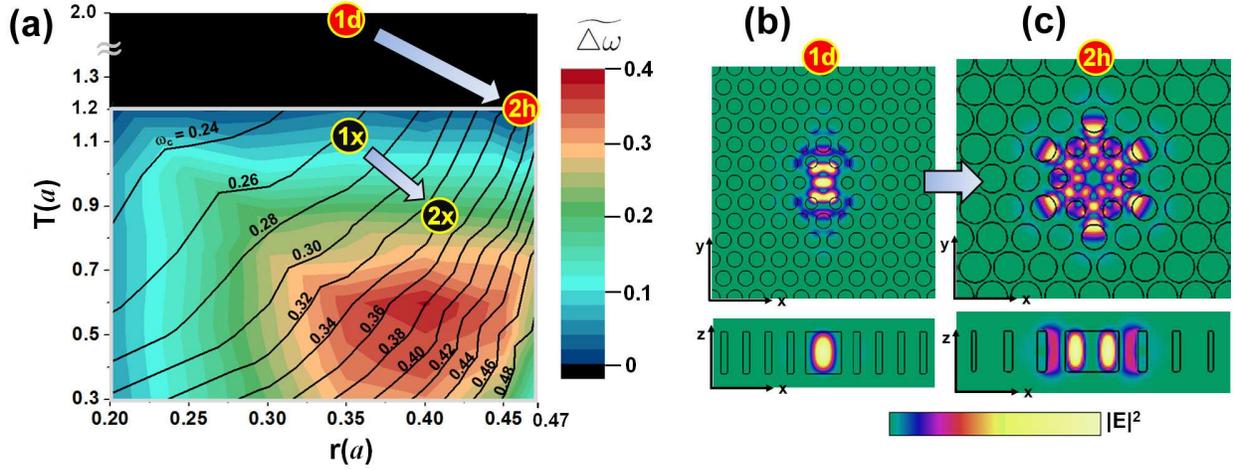}
\caption{\label{fig:fig1} (a) A 2-D map of a PBG for a triagular lattice air-hole (radius = $r$) PhC in a dielectric slab ($n_{\rm slab} = 3.4$)with a thickness of $T$. The 2-D color scale map represents the size of the PBG in terms of the gap-midgap ratio defined by $\widetilde{ \triangle\omega} \equiv \triangle \omega / \omega_c$, where $\omega_c$ is the center frequency of a PBG. The contour lines of $\omega_c$ are overlaid on the 2-D map. Note that throughout the Letter, all frequencies are normalized by $2 \pi c / a$, hence $\omega = a/\lambda$ (dimensionless). (b) The (first-order) dipole mode [$Q$=2,600 and $V=0.82(\lambda/n_{\rm slab})^3$] oscillating at $\lambda = 1341$ nm with $a$ = 325 nm and (c) the second-order hexapole mode [$Q$=15,200 and $V=2.23(\lambda/n_{\rm slab})^3$] oscillating at $\lambda = 1365$ nm with $a$ = 500 nm. Both modes are formed in a slab with $T$ = 606 nm.}
\end{figure*}

To start, we perform numerical simulations both using the plane-wave-expansion method (PWE)\cite{MPB} and the finite-difference time-domain method (FDTD) to investigate how a PBG evolves as we change the air-hole radius ($r$) and the slab thickness ($T$) [Fig.~\ref{fig:fig1}(a)]. Note that $r$ and $T$ are represented in the unit of the lattice constant ($a$). In the case of a triangular lattice air-hole PhC, $\{ r = 0.40 a, T = 0.6 a\}$ gives the widest gap centered at $\omega_c \approx 0.38$, which agrees with earlier work by Johnson, {\em et al.}\cite{JohnsonFa99}. Also note that there exists a broad region of $\{ r, T \}$ that gives a wide gap-to-midgap ratio\cite{JohnsonFa99} $\widetilde{ \triangle\omega}$ $>$ 30\%. This is why $r$ and $T$ are often chosen to be $\sim 0.35 a$ and $0.5 a$, respectively. We also find that a tiny PBG (usually $\widetilde{ \triangle\omega}$ $\sim$ 1\%) exists up to $T = 1.25 a$.

\begin{figure}[b]
\centering\includegraphics[width=8cm]{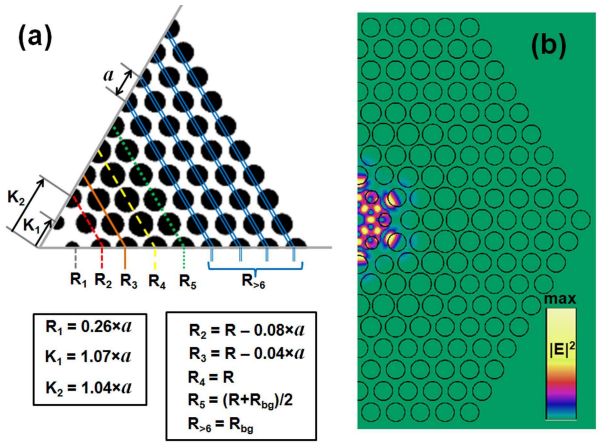}
\caption{\label{fig:fig2} (a) A schematic diagram shows how we finely tune air-hole sizes and locations to optimize $Q$. (b) An electric-field intensity distribution ($|{\bf E}|^2$) of the highest-$Q$ mode (case II in Table~\ref{table:table1}).}
\end{figure}

The dipole mode discussed in our previous work\cite{S_H_Kim_OL12} ($T = 1.86 a$) is marked as `1d' in the gap map. Now, we pose the question of {\em whether we can design a certain resonant mode emitting at $\sim$1.3 $\mu$m that is confined by a PBG in a slab with $T$ = 606 nm.} From the gap map diagram, the only possibility appears to be increasing $a$ in order to bring down $T(a)$ below $1.25 a$. However, keeping the same `1d' mode, larger $a$ usually results in the longer $\lambda$, because $\omega = a/\lambda$ is rather fixed by the in-plane modal structure of a resonant mode\cite{Joannopoulos_book}. Therefore, we should look instead into other resonant modes that do not resemble the dipole mode. 

It is well known that even a single defect resonator supports multiple resonances such as the quadrupole, the hexapole, and the monopole modes\cite{H_G_Park_QE02}. These higher-order modes are pulled down from the conduction band-edge of the photonic band structure\cite{Joannopoulos_book}. Further tuning the defect region can get more higher-order modes pulled down into the gap. 
One possible route from the (first-order) dipole mode (`1d') to the second-order hexapole mode (`2h') is drawn by an arrow in the gap map diagram. The `2h' is designed to be resonant at a wavelength close to that of `1d'\cite{2h_parameter} even though it has quite a large $a$ of 500 nm (thus, $T \approx 1.21 a$). As a quantitative measure showing how well the PhC layers work as a mirror, we calculate the vertical extraction efficiency $\eta_{\rm vert}$ defined by
$\eta_{\rm vert} \equiv ({1/Q_{\rm vert}})/(1/Q_{\rm horz} + 1/Q_{\rm vert}) = ({1/Q_{\rm vert}})/({1/Q_{\rm tot}})$\cite{S_H_Kim_OL12}.
We find that $\eta_{\rm vert}$ of `2h' shown in Fig.~\ref{fig:fig1}(c) is 0.954 ($Q_{\rm horz} = 3.3 \times 10^5$) with the same number of air-hole barriers shown in Fig.~\ref{fig:fig2}(b). We believe this $\eta_{\rm vert}$ (or $Q_{\rm horz}$) has not yet been saturated due to the small gap size, expecting further improvement by increasing the number of barriers. 
Probably, in applying the idea of a higher-order resonant mode, $T$ of 606 nm would be the upper limit for $\lambda \sim$1300 nm, as `2h' can only be made barely located at the top-right corner of the gap map diagram. We would like to note that the same strategy can be applied more effectively to the case of an intermediate thickness range of 400 nm $< T <$ 600 nm. Imagine a first-order resonant mode (`1x') oscillating at $\omega \approx 0.26$ within a slab with $T = 1.1 a$. At this region, $\widetilde{ \triangle\omega}$ is only about 5\%. We can bring it down deep into the band gap by utilizing its second-order resonant mode (`2x'). If `2x' oscillates at $\omega \approx 0.33$, then, without altering $\lambda$, `2x' can be formed in a slab with $T \approx 0.87 a$, at which $\widetilde{ \triangle\omega}$ is as large as 20\%. 

\begin{table}[b]
  \centering
  \caption{\bf \label{table:table1} Examples of the second-order hexapole mode in a $T$ = 606 nm slab.}
   \begin{tabular}{cccccc} \\ \hline
    case & $R (a)$ & $R_{\rm bg} (a)$ & $Q_{\rm tot}$ & $Q_{\rm vert}$ & $\eta_{\rm vert}$ \\ \hline
    {\rm I} & 0.45 & 0.45 & 55,400 &  58,500 & 0.947 \\
    {\rm II} & 0.45 & 0.38 & 105,100 & 146,200 & 0.719 \\
    {\rm III} & 0.44 & 0.38 & 50,400 & 63,900 & 0.789 \\
    {\rm IV} & 0.43 & 0.38 & 27,900 & 34,400 & 0.811 \\
    {\rm V} & 0.42 & 0.38 & 17,800 & 21,800 & 0.813 \\
    {\rm VI} & 0.41 & 0.38 & 12,400 & 15,300 & 0.807 \\ \hline
  \end{tabular}
\end{table}


Surrounding the `2h' with the large air-holes of $R = 0.46 a$ would give better spectral matching between the `2h' resonance and the center of the tiny bandgap. However, such a large air-hole radius is not advantageous for the device's mechanical robustness.  
Therefore, we proceed to study if the background air-hole radii ($R_{\rm bg}$) can be substantially reduced without sacrificing $Q$ too much. Several representative cases of fine-tuned air-holes are shown in Fig.~\ref{fig:fig2} and Table~\ref{table:table1}. $\omega$ and $Q$ are most sensitively dependent on the parameters near the center of the resonator; $R_1$, $K_1$, and $K_2$. These parameters have been determined in a manner to optimize $Q$. The out-skirt region from $R_4$ is intended as a mirror. Air-hole radii before and after $R_4$ are designed to vary gradually to minimize unintentional scattering losses at the crystal dislocations. Since $R_1$, $K_1$, and $K_2$ are fixed, all the resonant wavelengths tend to stay near 1323 nm. $a$ = 450 nm for all those cases, thus $T = 1.35 a$ and {\em there exists no PBG}. 

Contrary to the initial expectation, {\em $Q$ can be made higher even in the absence of a rigorous PBG}\cite{1st_hexa_thick}. In Case II, we find that $Q_{\rm vert}$ can be greatly improved by more than a factor of 10, thereby $Q_{\rm tot}$ can reach over $10^5$. It is interesting to observe that, comparing I and II, air-holes located far from the mode's energy ($R_{>6}$) can affect $Q_{\rm vert}$. It should also be noted that just one layer of $R_4$ = $0.45 a$ effectively blocks the horizontal photon leakage. As we progressively reduce $R$, both $Q_{\rm vert}$ and $Q_{\rm horz}$ decrease somewhat. At the final stage of the tuning (VI), all air-hole sizes become reasonable for experimental realization and $Q$ remains well above what is required for lasing. 

\begin{figure}[t]
\centering\includegraphics[width=8.5cm]{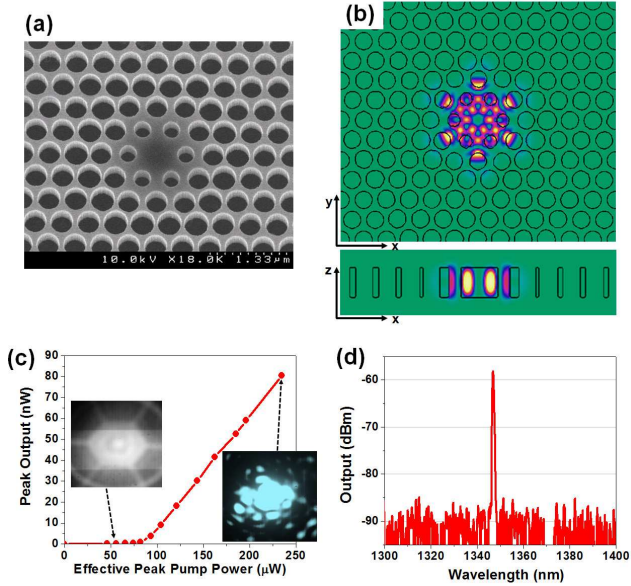}
\caption{\label{fig:fig3} (a) Scanning electron microscope (SEM) image taken at a tilt of about 10$^{\circ}$. Note that $T$ = 606 nm. (b) FDTD simulated mode profile. The actual SEM image was used to directly input air-hole shapes and sizes. (c) Light-in versus light-out ({\em L-L}) curve. Insets show near-infrared camera images taken before and after the lasing. (d) Lasing spectrum measured at a peak pump power of 200 $\mu$W.}
\end{figure}

In experiment, we intend to fabricate structurally more robust design similar to VI rather than the $Q$-optimized design of II. We use the same InGaAsP wafer containing 7 InGaAsP quantum wells emitting near 1325 nm used in our previous work\cite{S_H_Kim_OL12}. To define high-aspect ratio air-holes, we use chemically-assisted ion-beam etching with Ar and ${\rm Cl}_2$\cite{S_H_Kim_OL12}. 
The fabricated devices are optically pumped at room-temperature with a 830 nm laser diode driven by a pulse generator at 1 MHz with a duty cycle of 2.5 \%. A 100$\times$ objective lens is used to focus the pump laser on the center of the resonator. The {\em L-L} curve clearly shows a threshold, estimated to be 88 $\mu$W in terms of peak pump power, where we have assumed about 20\% of actual incident pump power is absorbed by the slab. We verify single mode lasing operation over a wide spectral range (1300 nm$\sim$1400 nm) with a side-mode suppression ratio of $\sim$30 dB. To confirm if the measured laser peak truly originates from the `2h', we perform FDTD simulation using a contour input for actual fabricated air-holes from the SEM image. The FDTD expects that the designed `2h' mode should locate at a wavelength of 1340 nm, which agrees very well with the experimental result. 

In summary, we show that a PhC slab with optically thick $T$ = 606 nm can be used to construct a PBG-confined resonant mode oscillating at a wavelength of $\sim$1300 nm. We also show that a surprisingly high $Q$ of over $10^5$ can be obtained even in the absence of a rigorous PBG, and that the use of the higher-order resonant mode can be quite advantageous for making an efficient PhC laser with an optically thick slab. 

\bigskip

This work was supported by the Defense Advanced Research Projects Agency under the Nanoscale Architecture for Coherent Hyperoptical Sources program (W911NF-07-1-0277) and by the National Science Foundation under NSF CIAN ERC (EEC-0812072).

\bibliographystyle{osajnl}
\bibliography{HighQThickSlabA}

\begin{thebibliography}{10}
\bibitem{2h_parameter}
Structural parameters for this 2nd hexapole mode are as follows: $K_1 = 1.07
  a$, $K_2 = 0.99 a$, $R_1 = 0.28 a$, $R_{\rm bg} = 0.46 a$, $R = 0.46 a$. For
  definitions of these parameters, see Fig. ~\ref{fig:fig2}.
\end{thebibliography}

\end{document}